\documentclass[twocolumn,pre,superscriptaddress]{revtex4}
\usepackage{graphicx,float}
\usepackage{latexsym}
\usepackage{epstopdf}
\usepackage{amssymb,amsmath}
\usepackage{color}
\usepackage{multirow}
\usepackage{hyperref}
\usepackage{graphicx}
\makeatletter
\let\Hy@linktoc\Hy@linktoc@none
\makeatother

\hypersetup{colorlinks=true,linkcolor=blue,citecolor=red}
\usepackage{verbatim}
\usepackage{bm}

\newcommand*\Laplace{\mathop{}\!\mathbin\bigtriangleup}
\begin{document}

\title{Modeling and statistical analysis of non-Gaussian random fields with heavy-tailed distributions}

\author{Mohsen Ghasemi Nezhadhaghighi }
\email{ghaseminejad@shirazu.ac.ir}
\affiliation{Department of Physics, Shiraz University, Shiraz  71454, Iran}
\author{Abbas Nakhlband}
\affiliation{Department of Physics, Sharif University of Technology, Tehran 14588-89694, Iran}

\begin{abstract}
In this paper, we investigate and develop a new approach to the numerical analysis and characterization of random fluctuations with heavy-tailed probability distribution function (PDF), such as turbulent heat flow and solar flare fluctuations. We identify the heavy-tailed random fluctuations based on the scaling properties of the tail exponent of the PDF, power-law growth of $q$th order correlation function and the self-similar properties of the contour lines in two-dimensional random fields. Moreover, this work leads to a substitution for fractional Edwards-Wilkinson (EW) equation that works in presence of $\mu$-stable L\'evy noise. Our proposed model explains the configuration dynamics of the systems with heavy-tailed correlated random fluctuations. We also present an alternative solution to the fractional EW equation in the presence of $\mu$-stable L\'evy noise in the steady-state, which is implemented numerically, using the $\mu$-stable fractional L\'evy motion. Based on the
analysis of the self-similar properties of contour loops, we numerically show that the scaling properties of contour loop ensembles can qualitatively and quantitatively distinguish non-Gaussian random fields from Gaussian random fluctuations.
\end{abstract}

\maketitle

\section{Introduction}\label{section 1}
The ubiquitous existence of Gaussian random fluctuations in natural phenomena has been subject of plenty of interest in many fields, such as condensed matter physics, cosmology, string theory, biology, economics and earth science \cite{mandelbrot,surface2,sornette,mandelbrot1,falconer,nelsonbook,adler}. Experimental measurements and numerical simulations 
show the presence of particular scaling properties in time and space in the fluctuations of Gaussian random fields. For
instance, growing interfaces and elastic membranes were known to be described by a finite set
of geometrical scaling exponents \cite{surface2,nelsonbook}.

In Refs. \cite{kondevprl,kondevpre} the self-similar properties of two-dimensional Gaussian random fields are discussed by the analysis of the scaling properties of \textit{contour loop ensembles}. Using the scaling analysis of contour loops, it has been shown that such properties are valid for the contour lines in glassy interfaces and turbulence \cite{kondevother}, two-dimensional fractional Brownian motion \cite{rajabpour2009}, discrete scale-invariant rough
surfaces \cite{haghighi2011}, synthetic multifractal rough surfaces \cite{hosseinabadi2012}, discrete surface growth models in (2+1)Dimensions \cite{hosseinabadi2014}, isothermal loops on the cosmic microwave background \cite{cmbcontours}, and electron-hole puddles in Graphene \cite{sarma2011,najafi2016}. 
In most cases, fast decaying distributions are a noteworthy characteristic of the above mentioned examples. Thus, they seem, not to be significantly different from Gaussian random fluctuations \cite{kondevprl,kondevpre}. 
However, in reality, there are many cases where nonlinear effects result in strong correlations in the system. In addition, in these systems, the probability distributions often
display heavy-tailed non-Gaussian behavior \cite{sornette}.

There have been many studies exploring heavy-tailed non-Gaussian behaviors of the probability density function of natural phenomena, such as fluctuations in the resistance of conducting wires \cite{weissman1988,Orlyanchik2006}, the statistics of turbulent fields \cite{sorriso,biskamp} and turbulent heat flows \cite{kadanoff}, the global velocity of imbibition fronts \cite{planet2009}, density of a vibrated column of granular material \cite{Nowak}, solar flare fluctuations \cite{Scafetta}, particle velocity fluctuations in granular systems \cite{Radjai}, and the intensity histograms of individual images \cite{Ruderman}

Our focus, in this paper, is modeling, simulation and characterization of the random fluctuations with power-law tails in their probability density functions. In order to drive some statistical properties of random fluctuations with heavy-tailed PDF, we will begin with proposing a rather simple linear differential equation \textit{i.e.} fractional Edwards-Wilkinson equation in the presence of $\mu$-stable L\'evy noise. The scaling theory implies the existence of a few scaling laws and exponents. For example, $q$th order two-point correlation function has power-law behavior with exponent $\alpha$, the so-called \textit{roughness exponent}. In order to further characterize the random fluctuations with heavy-tailed distribution, the scaling properties of our proposed model are computed in the steady state, using the $\mu$-stable fractional L\'evy motion. In particular we will examine the power-law behavior of the $q$th order correlation functions and self-affine properties of the contour loop ensembles. We will show that there are some ``non-negligible'' differences between the scaling properties of Gaussian random fluctuations and heavy-tailed non-Gaussian random fluctuations in two spatial dimensions. 

The rest of this paper is organized as follows: In the Section \ref{section 2}, we will introduce a linear dynamical process that describes the out-of-equilibrium interface dynamics with Gaussian statistics. Then a computational model for simulating random Gaussian process, such as fractional Brownian motion, is presented. In the Section \ref{section 3}, a substitution for the fractional Edwards-Wilkinson equation is introduced. This equation works in the presence of $\mu$-stable L\'evy noise for modeling of the systems with heavy-tailed non-Gaussian PDF. Also, we have studied particular scaling properties of our proposed model. In the same section, we provide a general method which generates random fluctuations with heavy-tailed PDF. In the Section \ref{section 4}, simulations of $\mu$-stable fractional L\'evy motion and also an analysis based on the scaling properties of the probability distributions are discussed. Finally, the summary and the conclusions are presented in Section \ref{section 5}.

\section{Random fluctuations with Gaussian statistics}\label{section 2}
\subsection*{A linear model for Gaussian random fluctuations}
The simplest stochastic partial differential equations that describes the out-of-equilibrium interface
dynamics with Gaussian statistics, are the Edwards-Wilkinson (EW) and noisy Mullins-Herring (MH) equations  \cite{surface2,nelsonbook}. The usual generalized form of EW and MH equations can be written as
\begin{eqnarray}\label{ew mh}
\frac{\partial h(\vec{x},t)}{\partial t} = -D_m  (-\nabla^2)^{m/2} h(\vec{x},t) + \eta(\vec{x},t),
\end{eqnarray}
where $m$ is an \textit{even} number. For $m = 2$ we recover the EW equation, whereas $m = 4$ yields the MH equation. Here $h(\vec{x},t)$ is the scalar field parameterized by space coordinates $\vec{x}$ and time $t$. The coefficient $D$ is a constant and $\eta(\vec{x},t)$ is usually assumed to be a Gaussian white noise with $\langle \eta(\vec{x},t) \rangle = 0$ which satisfies the fluctuation-dissipation relation
 $\langle \eta(\vec{x},t)\eta(\vec{x}^\prime,t^\prime) \rangle = 2k_BT \delta(\vec{x}-\vec{x}^\prime)\delta(t-t^\prime)$ \cite{surface2,nelsonbook}. 
 
It is worth mentioning that, in the thermodynamic limit when $t\rightarrow \infty$ the central limit theorem for independent random variables asserts that the probability distribution function of the random profile $h(\vec{x},t)$ should converge to
a Gaussian form: $\mathcal{P}\lbrace h \rbrace \propto e^{-h^2/2\sigma_h}$ \cite{stochastic book,gardiner}.

Defining the Fourier-Laplace transform as
\begin{eqnarray}
h(\vec{k},s) = \int d^d x dt \exp(-s t - i\vec{k}.\vec{x}) h(\vec{x},t).
\end{eqnarray}
Under this transformation the stochastic partial differential equation (\ref{ew mh}) becomes:
\begin{eqnarray}
h(\vec{k},s) = \frac{\eta(\vec{k},s)}{s+D_m |k|^m},
\end{eqnarray}
where the Fourier transformed noise $\eta(\vec{k},t)$ has a zero mean, and its correlations obey the equation $\langle \eta(\vec{x},t)\eta(\vec{x}^\prime,t^\prime) \rangle = 2k_BT (2\pi)^d \delta^d (\vec{k}+\vec{k}^\prime)\delta(t-t^\prime)$. By inverting the Laplace transform, one can obtain 
\begin{eqnarray}
h(\vec{k},t) = \int_0^t e^{-D_m |k|^m(t-t^\prime)}\eta (\vec{k},t^\prime)dt^\prime.
\end{eqnarray}
The theory of kinetic roughening predicts that the random fluctuation $h(\vec{x},t)$ is correlated in space and time \cite{nelsonbook}. The simplest quantitative measure of a given random process is its spectral autocorrelation function, given by
\begin{eqnarray}
&&\langle h(\vec{k},t) h(\vec{k}^\prime,t) \rangle \equiv \nonumber \\ 
&& (2\pi)^d \delta^d (\vec{k}+\vec{k}^\prime) \frac{k_BT}{D_m|k|^m} \left[1-e^{-2D_m |k|^m t} \right].
\end{eqnarray} 
In the limit $t\to \infty$, we have the steady state structure function 
\begin{eqnarray}
S(\vec{k}) = \frac{k_BT}{D_m|k|^m}.
\end{eqnarray}
More generally, one can look at the matter of dealing with the linear stochastic partial differential equation (\ref{ew mh}) in such a way that when $m$ takes a fractional order $z$ we have: $(-\nabla^2)^{m/2}\mapsto (-\nabla^2)^{z/2}$. Introducing the fractional Langevin equation for random fields, which were studied prior to this work, it is shown that steady state structure function behaves as \cite{Masoudi2013}:
\begin{eqnarray}\label{spectral fractional ew}
S(\vec{k}) \propto \frac{1}{|k|^z},
\end{eqnarray}
where $z>0$ is the fractional order. 

Since the fractional Langevin approach provides an efficient method to study the statistical properties of the Gaussian random fields, one must consider the other possible methods  to generate correlated random fluctuations, such as fractional Brownian motion. Such methods describe the steady states of linear growth
models, \textit{i.e.} Eq. (\ref{ew mh}).

\subsection*{Fractional Brownian motion}

Fractional Brownian motion $\mathcal{B}(\vec{r})$, in ${\Re}^d$, is a Gaussian, non-stationary, and continuous random field. Its correlation function follows power-law scaling behavior described by \textit{Hurst} index $H$, while $0 \leq H \leq 1$. 
A wide range of physical processes can be modeled by fractional Brownian random field, including, biological, economical, geological phenomena \cite{peitgen}. 
Several methods have been commonly used to generate multidimensional fBm. 
The midpoint displacement method \cite{peitgen}, the successive random addition method \cite{peitgen}, the optimization method \cite{hamzehpour2006} and the Weierstrass-Mandelbrot function\cite{ausloos1985} are the most common methods to generate fBm samples. A  very efficient way to generate fractional Brownian random field is the modified Fourier filtering method \cite{makse1996}. To simulate Gaussian random fields in $d$-dimensions with this method, we consider the following power spectrum \cite{makse1996}:
\begin{eqnarray}\label{spectral fbm}
S(\vec{k})\equiv \langle |\tilde{B}(\vec{k})|^2 \rangle  = \frac{C_d}{\left( \sum_i k_i^2 + k_c^2 \right)^{H+d/2}},
\end{eqnarray}
in which $C_d$ is an arbitrary constant and $\vec{k}\equiv (k_1,\dots,k_d)$, with $k_i$ being the Fourier components in the $i$th direction. In Eq. (\ref{spectral fbm}), $k_c$ is the cutoff wave vector and $H$ is the \textit{Hurst} exponent. Two points belonging to the same random field, separated with
distance $r < 1/q_c$, are correlated, while this correlation diminishes
for $r > 1/q_c$ \cite{makse1996}. For the small values of $k_c$, Eq. (\ref{spectral fbm}) leads to the asymptotic power-law behavior: $S(\vec{k})\sim |\vec{k}|^{-(2H+d)}$. 
Note that the scaling form of the power spectrum $S(\vec{k})$ coincides with the spectral correlation function,  Eq. (\ref{spectral fractional ew}). Therefore, we expect a linear relation between fractional order $z$ in Eq. (\ref{spectral fractional ew}) and \textit{Hurst} index $H$:
\begin{eqnarray}
z=2H+d.
\end{eqnarray}
It is worthwhile mentioning that two-dimensional fractional Brownian random field ($d=2$), with $H = 0$,  corresponds to the Edwards-Wilkinson equation universality class ($z=2$) \cite{surface2}, and $H = 1$ describes the
universality class of the Mullins-Herring model ($z=4$) \cite{surface2}.

To simulate a random Gaussian surface in  two dimensions, one can consider a square lattice with sides of length $L$ and with grid size $a$.  The associated value for each cell $\mathcal{B}(x,y)$, where $(x,y)\equiv (ma,na)$, is given by Fourier-transforming $\tilde{\mathcal{B}}(k_x,k_y) $, where the Fourier components $k_x$ and $k_y$ take their values in the range $\left[-\pi/a,\pi/a\right]\times \left[ -\pi/a,\pi/a\right]$. To obtain the correlated surface with the desired correlation exponent $H$, one needs to calculate the inverse Fourier transform:
\begin{eqnarray}
\mathcal{B}(x,y) &\equiv & \mathcal{F} ^{-1}\left\lbrace \tilde{\mathcal{B}}(k_x,k_y)\right\rbrace \nonumber \\  
&=&  \mathcal{F}^{-1} \left\lbrace S^{1/2}(k_x,k_y)\eta (k_x,k_y)\right\rbrace,
\end{eqnarray}
where $S(k_x,k_y)  = {C_2}/{\left( k_x^2 +k_y^2 + k_c^2 \right)^{H+d/2}}$, and $\eta (k_x,k_y)$ is an independent uncorrelated Gaussian set of random variables associated with each possible values of $k_x$ and $k_y$ in the range of $\left[-\pi/a,\pi/a\right]\times \left[ -\pi/a,\pi/a\right]$. Thanks to the central limit theorem, the probability density function $\mathcal{P}\lbrace B(\vec{x}) \rbrace$ will be Gaussian.

\subsection*{Scale invariance in Gaussian random fluctuations}

The most interesting feature of the Langevin equation (\ref{ew mh}) is the scaling behavior of the interface fluctuations. The scaling properties of the random field $h(\vec{x},t)$ can be characterized in terms of the height-height correlation function by
\begin{eqnarray}\label{local roughness scaling}
C(r,t) \equiv \langle \left[ h(\vec{x},t)) - h(\vec{x}^\prime,0)\right]^{2} \rangle ^{1/2} \sim r^{\alpha}f(r t^{-1/\nu}),
\end{eqnarray}
where $r = |\vec{x}-\vec{x}^\prime|$, and the angular brackets denote the ensemble average. The scaling function $f(u)$ behaves like
\begin{eqnarray}
f(u) =
  \begin{cases}
    u^{-\alpha}       & \quad u\gg 1 \\
    \text{const.}  & \quad u \ll 1\\
  \end{cases}.
\end{eqnarray}
Another interesting measure, associated with fluctuating random fields in $d$-dimensional space at a given system
size $L$, is the global width 
\begin{eqnarray}\label{roughness measure}
W(L,t) \equiv \left\langle \int d^dx \left[h(\vec{x},t) - \bar{h}(t) \right]^2  \right\rangle^{1/2},
\end{eqnarray}	
where $\bar{h}(t) =\int d^dx h(\vec{x},t)$. In many cases, the global roughness is observed to satisfy the dynamic
scaling; such that
\begin{eqnarray}\label{global roughness scaling}
W(L,t) = 
\begin{cases}
    t^{\alpha_g/\nu}       & \quad t\ll t_\times \\
    L^{\alpha_g}  & \quad t \gg t_\times\\
  \end{cases},
\end{eqnarray}
where $t_\times = L^\nu$ and $\alpha_g$ is the global roughness exponent. The random fluctuation, when $\alpha_g = \alpha$, is self-affine \cite{surface2}. The ratio $\beta = \alpha_g/\nu$ in Eq. (\ref{global roughness scaling}) is called the time exponent. 

The scaling properties of the fractional Edwards-Wilkinson equation has been studied in Re. \cite{nezhadhaghighi2014}. Since this model is linear the scaling exponents $\alpha$, $\beta$ and $\nu$ can be easily calculated: 
\begin{eqnarray}
\alpha = \frac{z-d}{2},~ \beta = \frac{z-d}{2z}, ~ \nu = z.
\end{eqnarray}

We stress that the fractional Brownian random field resulted from the power spectrum $S(\vec{k})\sim |\vec{k}|^{-(2H+d)}$ is also scale invariant. The height correlation function, which is related to the height power spectrum, has the scaling form $C(r)\equiv \int  S^{1/2}(\vec{k})(e^{i\vec{k}.\vec{x}}-1)d^d \vec{k} \propto r^{H}$. From this scaling form we conclude that there is a one-to-one correspondence between the stationary solution of the fractional version of the stochastic differential equation (\ref{ew mh}) and the fractional Brownian motion with $H = \frac{z-d}{2}$.

\section{Random fluctuations with power-law statistics}\label{section 3}

\subsection*{A linear model for heavy-tailed random fluctuations}
\begin{figure}[t]
\begin{center}

\includegraphics[width=8cm,clip]{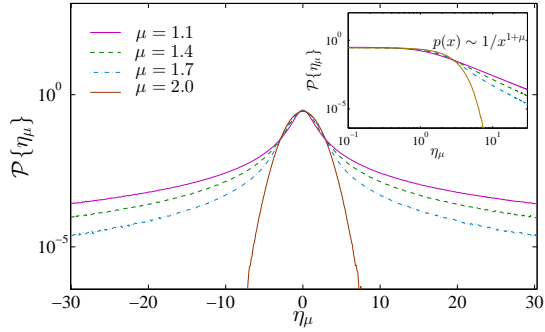}
\end{center}
\caption{(Color online) The probability distribution function $\mathcal{P}\lbrace \eta_\mu \rbrace$ for the symmetrical $\mu$-stable L\'evy process in semi-log scale. (Inset) the same figure in $\log$-$\log$ scale. The special case $\mu = 2.0$ corresponds to the Gaussian distribution function.}
\label{figur1}
\end{figure}

Here we are interested in modifying the fractional generalization of the Eq. (\ref{ew mh}) in order to investigate and quantify the statistical properties of non-Gaussian heavy-tailed random field. In this paper, we consider the following stochastic partial differential equation
\begin{eqnarray}\label{fractional edwards wilkinson}
\frac{\partial h(\vec{x},t)}{\partial t} = D_z   \frac{\partial ^z}{\partial |\vec{x}|^z}h(\vec{x},t) + \eta_{\mu}(\vec{x},t) 
\end{eqnarray}
where $\frac{\partial ^z}{\partial |\vec{x}|^z}$  is the multidimensional Riesz-Feller fractional space derivative of order $z$, and $\eta_{\mu}(\vec{x},t)$ is $\mu$-stable L\'evy noise $\eta_{\mu}(\vec{x},t)$ with $\langle \eta_{\mu}(\vec{x},t) \rangle = 0$. The parameter $\mu >0$ characterizes the asymptotic power-law behavior (see Fig.(\ref{figur1})) of the stable distribution for the $\mu$-stable L\'evy noise:
\begin{eqnarray}\label{power law PDF}
p(x) \sim \frac{1}{|x|^{1+\mu}}.
\end{eqnarray}

Riesz-Feller fractional operator is also defined via its Fourier transform through the functional relation 
\begin{eqnarray}
\mathcal{F}\left\lbrace \frac{\partial ^z}{\partial |\vec{x}|^z} h(\vec{x},t) \right\rbrace \equiv -|\mathbf{q}|^z h(\mathbf{q},t),
\end{eqnarray}
where $h(\mathbf{q},t)$ is Fourier transform of the random field $h(\vec{x},t)$.
It is also possible to rewrite the Riesz-Feller derivative using the standard Laplacian $\Laplace$ as $\frac{\partial ^z}{\partial |\vec{x}|^z}\equiv -(\Laplace)^{z/2}$\cite{samko,kilbas}. 

We should note that the noise term in the Eq. (\ref{fractional edwards wilkinson}) with $\mu = 2$ corresponds to the usual uncorrelated Gaussian white noise. Besides, $z=2$ and $z=4$ correspond to the ordinary Edwards-Wilkinson and Mullins-Herring equations (\ref{ew mh}). Moreover, with $0 < \mu < 2$, the corresponding L\'evy white noise has infinite variance and the higher cumulants \cite{ps1}. For infinite variance models, the generalized central limit theorem states that the average of a large number of independent identically distributed random variables with the probability distribution Eq. (\ref{power law PDF}) converges to the heavy-tailed stable distribution. 

The solution to Eq. (\ref{fractional edwards wilkinson}) is a scale invariant function. Therefore, we have
 \begin{eqnarray}
h(\lambda \vec{x},\lambda^\nu t) = f(\lambda) h(\vec{x},t),
 \end{eqnarray}
where $f(\lambda)$ is a power function of the scale factor $\lambda$. In the following, we will analyze the scaling properties of the non-Gaussian random fields generated by the solution of Eq. (\ref{fractional edwards wilkinson}), in and out of equilibrium. 

Consider the fractional Langevin equation (\ref{fractional edwards wilkinson}) under a scaling transformation $\vec{x}\rightarrow \lambda \vec{x}$ and $t\rightarrow \lambda^\nu t$, together with a specified rescaling of the fields $h\rightarrow \lambda^\alpha h$. Using the scale transformations, the Eq. (\ref{fractional edwards wilkinson}) yields
\begin{eqnarray}\label{scale transform fractional edwards wilkinson}
\lambda^{\alpha - \nu}\frac{\partial h(\vec{x},t)}{\partial t} = \lambda^{\alpha - z} D_z   \frac{\partial ^z}{\partial |\vec{x}|^z}h(\vec{x},t) + \lambda^ {\gamma} \eta_{\mu}(\vec{x},t), 
\end{eqnarray}
where $\gamma = (d+z)(1/\mu -1)$, and $d$ is the space dimension. For the Eq. (\ref{fractional edwards wilkinson}) to be scale invariant it is required that 
\begin{eqnarray}\label{scaling exponent for fEW}
\nu = z \textrm{ and } \alpha = \frac{z+(1-\mu)d}{\mu},
\end{eqnarray}
where $\nu$ and $\alpha$ are called the growth and the \textit{roughness} exponents, respectively.

In the next section, we will introduce $\mu$-stable fractional L\'evy motion, which provides a perfect candidate for the solution to Eq. (\ref{fractional edwards wilkinson}) in the saturated regime (when $t\rightarrow \infty$).

\subsection*{$\mu$-stable fractional L\'evy motion}

The most well-known examples of the stochastic random fields with Gaussian distributed increments is the $d$-dimensional fractional Brownian motion (fBm). In contrast to the usual fBm with Gaussian statistics, we will propose a new algorithm, $\mu$-stable fractional L\'evy motion ($\mu$fLm), to simulate random fields with non-Gaussian statistics. 
\begin{figure}[t]
\begin{center}

\includegraphics[width=8cm,clip]{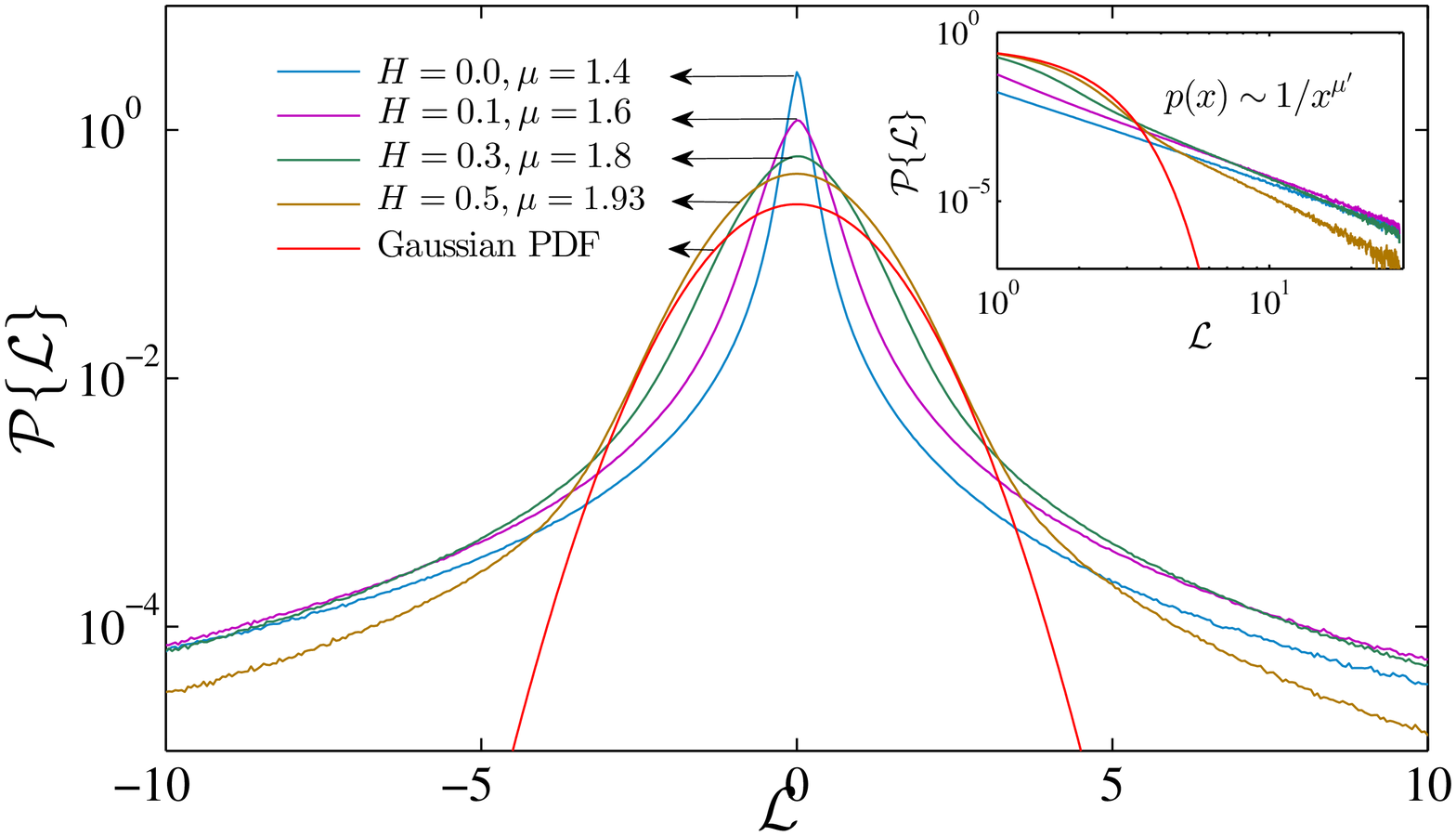}
\end{center}
\caption{(Color online) The probability distribution function $\mathcal{P}\lbrace \mathcal{L} \rbrace$ for the $\mu$-stable fractional L\'evy motion $\mathcal{L}(\vec{x})$ in semi-log scale. (Inset) the same figure in $\log$-$\log$ scale shows the scaling behavior $p(x) \sim 1/x^{\mu^\prime}$, where $\mu^\prime$ is the tail scaling exponent for the probability distribution function $\mathcal{P}\lbrace \mathcal{L} \rbrace$.}
\label{figur2}
\end{figure}
According to the generalized central limit theorem \cite{gardiner}, it would be possible to generate random fields with heavy-tailed fluctuations and non-Gaussian distribution function, \textit{i.e.} stable distributions. The best possible way to generalize fractional Brownian motion is to change the probability distribution of the random variables in the Fourier space. Therefore, we used a standard procedure to generate random variables $\eta_{\mu}(q_x,q_y)$ from a $\mu$-stable L\'evy distribution. One can generate the $\mu$-stable fractional L\'evy motion, $\mathcal{L}(\vec{x})$, over a substrate with a size $L\times L$ by the equation

\begin{eqnarray}\label{mu stable levy motion}
\mathcal{L}(x,y) \equiv  \mathcal{F}^{-1} \left\lbrace S^{1/2}(q_x,q_y)\eta_\mu (q_x,q_y)\right\rbrace,
\end{eqnarray}
where $\eta_{\mu}$s are independent and identically distributed random variables, and $S(\vec{q})=\frac{C_d}{\left( \sum_i k_i^2 + k_c^2 \right)^{H+d/2}}$. Let $p(\eta_{\mu})$ is the distribution function of the random variable $\eta_{\mu}$ with the asymptotic power-law behavior $p(x) \sim 1/x^{1+\mu}$.   
The probability density of a symmetrical $\mu$-stable L\'evy process is given by

\begin{eqnarray}\label{levy pdf}
p(x) = \frac{1}{\pi} \int _0^\infty e^{-\gamma k^{\mu}}\cos(kx)dk,
\end{eqnarray}
where $\gamma$ is the scale unit of the process, and $0<\mu\leq 2$ characterizes the distribution \cite{weron}. The spacial case $\mu=2$ corresponds to Gaussian PDF. The procedure of generating uncorrelated random variable $\eta_\mu$ corresponding to the probability distribution function, Eq. (\ref{levy pdf}), is 
\begin{eqnarray}
\eta_\mu = \frac{\sin(\mu V)}{\cos(V) ^{1/\mu}} \left[\frac{\cos((1-\mu)V)}{ W} \right] ^ {(1-\mu)/\mu},
\end{eqnarray}
where $V$ is a uniformly distributed random variable in the interval $(-\pi/2,\pi/2)$, and the random number $W$ has an exponential distribution with mean $1$ \cite{weron}.

The advantage of this algorithm for generating non-Gaussian, as well as Gaussian, random fields is demonstrated in the next section, where it is shown that the random profiles in two dimensions have the desired statistical properties.

For further analysis of the random fields with infinite variance fluctuations and heavy-tailed distributions, we need to study the scaling properties of the $q$th order moments (see comment following \cite{ps1}). The most useful measures for characterization of the heavy-tailed fluctuations are the $q$th order two-point correlation function,
\begin{eqnarray}\label{local roughness scaling}
C_q(r) \equiv \langle \left[ h(x)) - h(x^\prime)\right]^{q} \rangle ^{1/q} \sim r^{\alpha_l},
\end{eqnarray}
and the $q$th order \textit{global} width of the random field $h(x)$ defined by
\begin{eqnarray}\label{roughness measure}
W_q(L) \equiv \left\langle \overline{|h(x)-\overline{h(x)}|^q} \right\rangle ^{1/q} \sim L^{\alpha_g},
\end{eqnarray}	
where the overline denotes a spatial average over the system of size $L$, and the bracket denotes an ensemble averaging. Note that the $q$th order moments (\ref{local roughness scaling}) and (\ref{roughness measure}) are finite only if $0 < q < \mu$.

Based on the scaling arguments made previously for heavy-tailed random fluctuations (see Eq. (\ref{scale transform fractional edwards wilkinson})), we believe that $\alpha_g=\alpha_l=\alpha$ ($\alpha = \frac{z+(1-\mu)d}{\mu}$). 

Here, we mainly focus on the so-called $\mu$-stable fractional L\'evy motion, as an examples of the random fluctuations with heavy-tailed probability distribution functions. The two-dimensional $\mu$-stable fractional L\'evy fields also satisfy the scaling relations analogous to (\ref{local roughness scaling}) and (\ref{roughness measure}) with exponents $\alpha^\prime_l$ and $\alpha^\prime_g$. Our purpose is to numerically measure the local and global roughness exponents $\alpha^\prime_l$ and $\alpha^\prime_g$ as functions of the control parameters $H$ and $\mu$. We believe that for a given value of the stable parameter $\mu$, there is a direct relation between $\alpha^\prime$ and $\alpha$, which means that there is a linear relation between the fractional order $z$ in Eq. (\ref{fractional edwards wilkinson}) and the \textit{Hurst} index $H$ in $\mu$-stable fractional L\'evy motion. 

\section{Results}\label{section 4}

\begin{figure}[t]
\begin{center}

\includegraphics[width=8cm,clip]{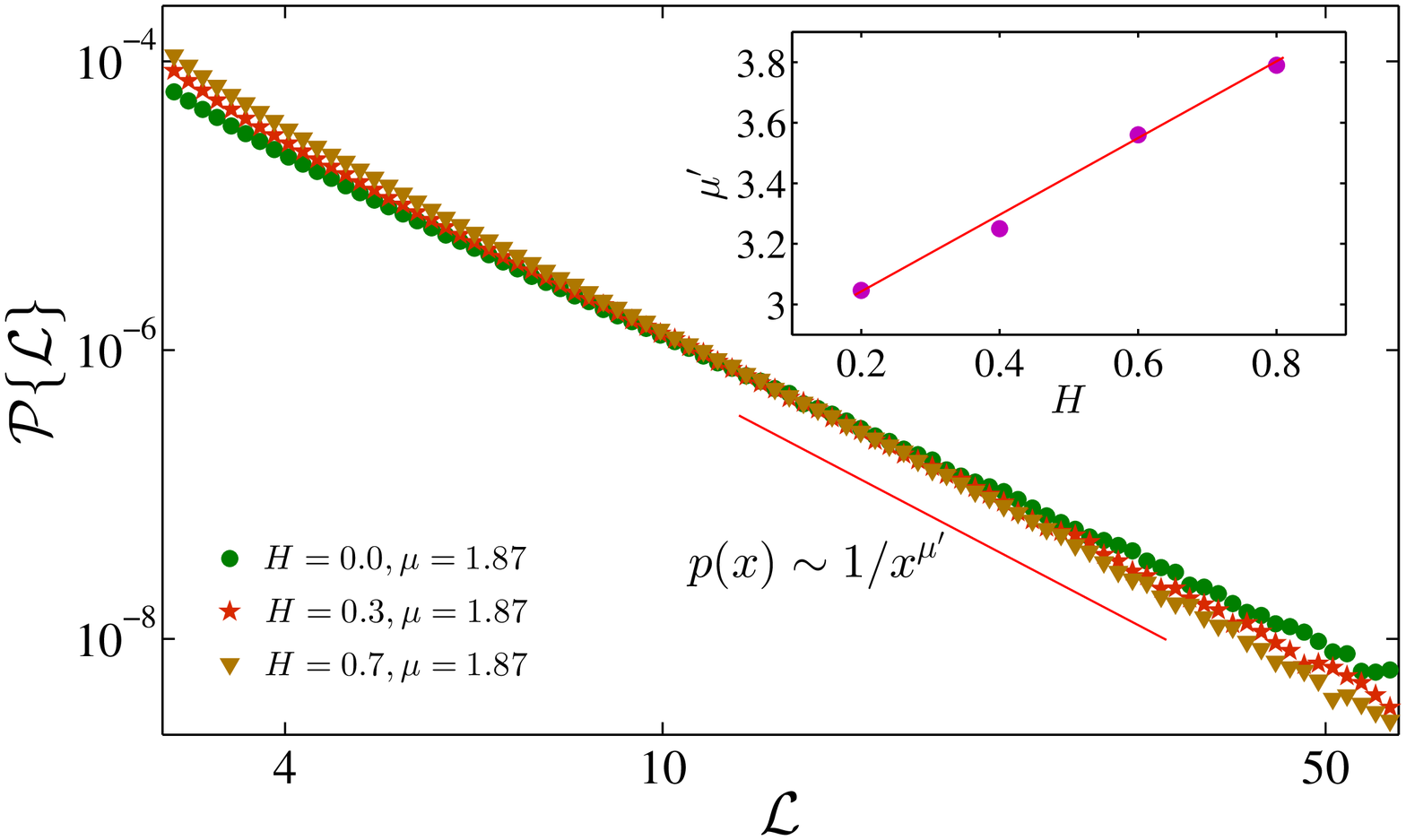}
\includegraphics[width=8cm,clip]{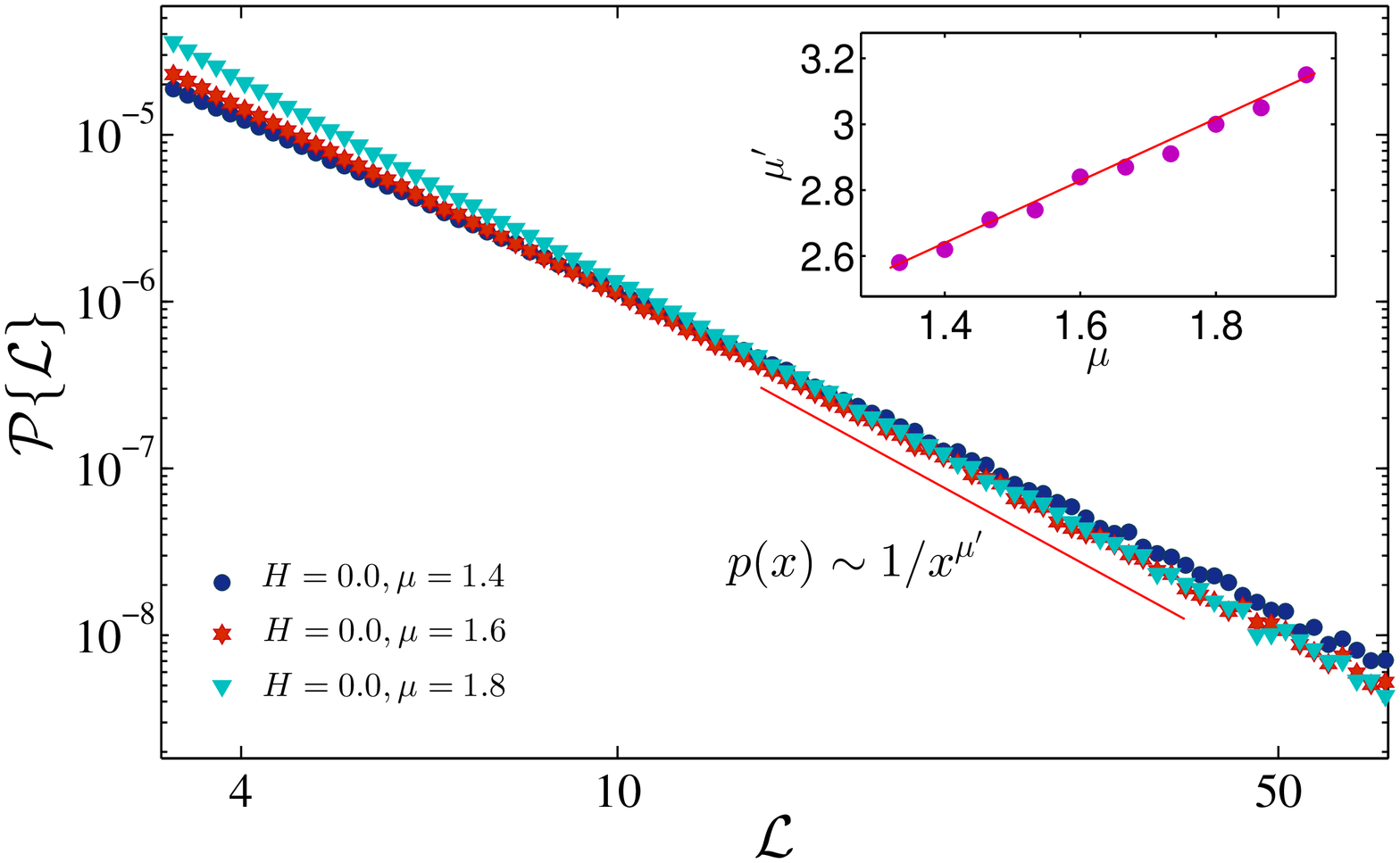}
\end{center}
\caption{(Color online) The probability distribution function $\mathcal{P}\lbrace \mathcal{L} \rbrace$ for the $\mu$-stable fractional L\'evy motion $\mathcal{L}(\vec{x})$ in $\log$-$\log$ scale, which shows the scaling behavior $p(x) \sim 1/x^{\mu^\prime}$ where $\mu^\prime$ is the tail scaling exponent for the probability distribution function $\mathcal{P}\lbrace \mathcal{L} \rbrace$. Top: The PDF for $\mu$-fLm with fixed value of $\mu = 1.87$ and different values of the \textit{Hurst} index $H$. The scaling exponent $\mu^\prime$  is expected to be closely related to the original \textit{Hurst} exponent and would be a linear function of $H$ (Inset). Bottom: The same analysis for $\mu$-fLm with fixed value of the \textit{Hurst} index $H=0.0$ and different values of $\mu$ which shows the linear dependence $\mu^\prime$ to the original value $\mu$ (Inset). }
\label{figur3}
\end{figure}

\subsection{Testing for Gaussianity and self-affinity} 
In this subsection, we will outline the essential properties of the non-Gaussian self-affine random fields. Here we considered the $\mu$-stable fractional L\'evy motion, Eq. (\ref{mu stable levy motion}) which is equivalent to the solution of the Eq. (\ref{fractional edwards wilkinson}) in the saturation limit.

\subsubsection{Probability density of non-Gaussian random fields}

Consider a single valued Gaussian random field $h(\vec{x})$ which satisfies up/down symmetry $h(\vec{x})\longleftrightarrow  -h(\vec{x})$. The most essential condition for the probability density of a Gaussian random field which is necessary to be satisfied is
\begin{eqnarray}\label{Gaussian fields PDF}
\mathcal{P}\left\lbrace h \right\rbrace \equiv \frac{1}{\sigma\sqrt{2\pi}}e^{-\frac{h^2}{2\sigma^2}},
\end{eqnarray}
where $\sigma^2$ is the variance of the fluctuation. 

As illustrated in the Fig. (\ref{figur2}), the probability distribution function of the $\mu$-stable fractional L\'evy motion $\mathcal{L}(\vec{x})$ with $\mu <2$ exhibits non-Gaussian behaviors with heavy-tailed distribution function. Conversely, for the limiting case $\mu \rightarrow 2$ the probability distribution function, $\mathcal{P}\lbrace \mathcal{L} \rbrace$, tends to Gaussian distribution (\ref{Gaussian fields PDF}) (see Fig. (\ref{figur2})). 

\begin{figure}[t]
\begin{center}
\includegraphics[width=8cm,clip]{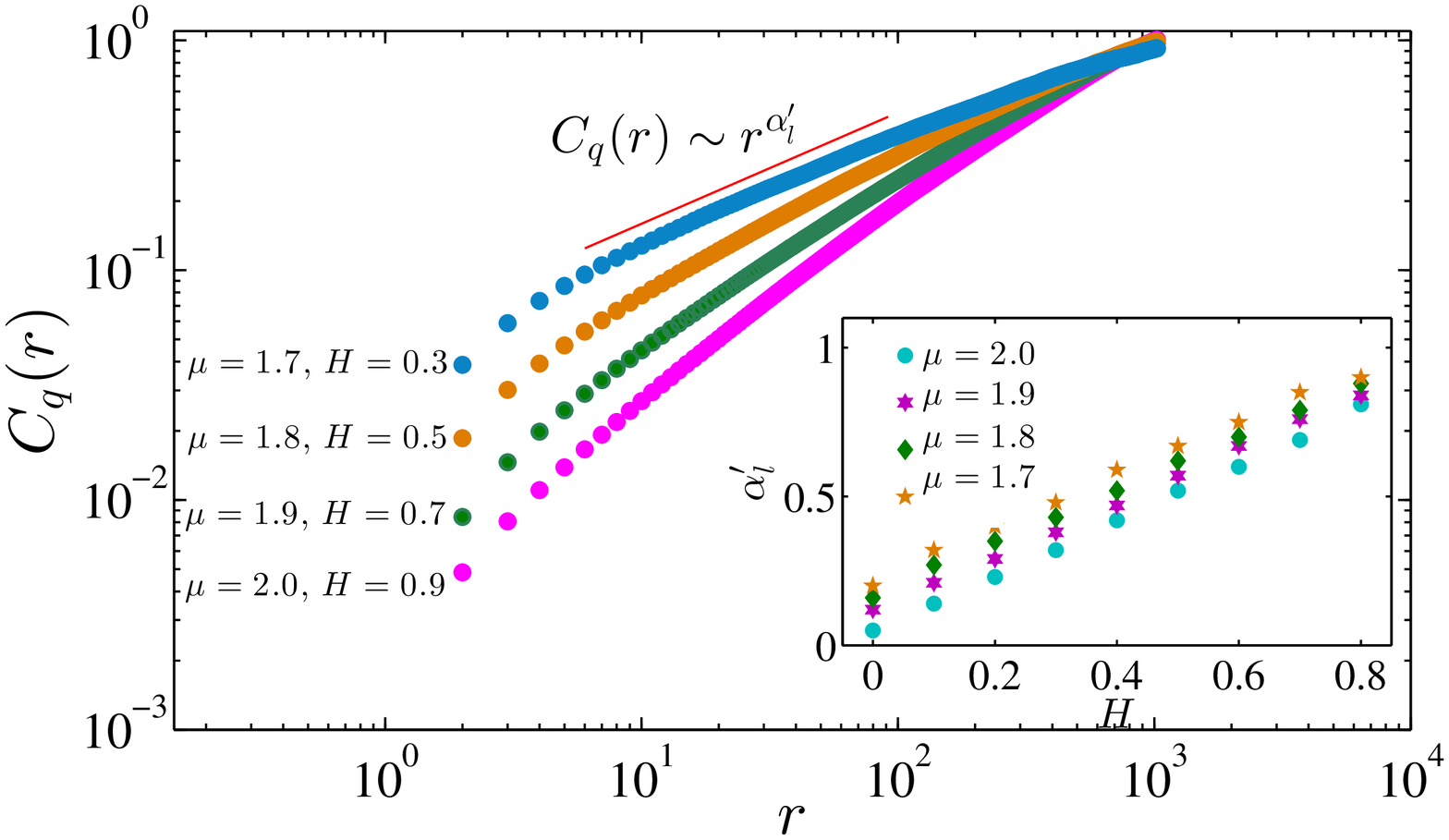}
\includegraphics[width=8cm,clip]{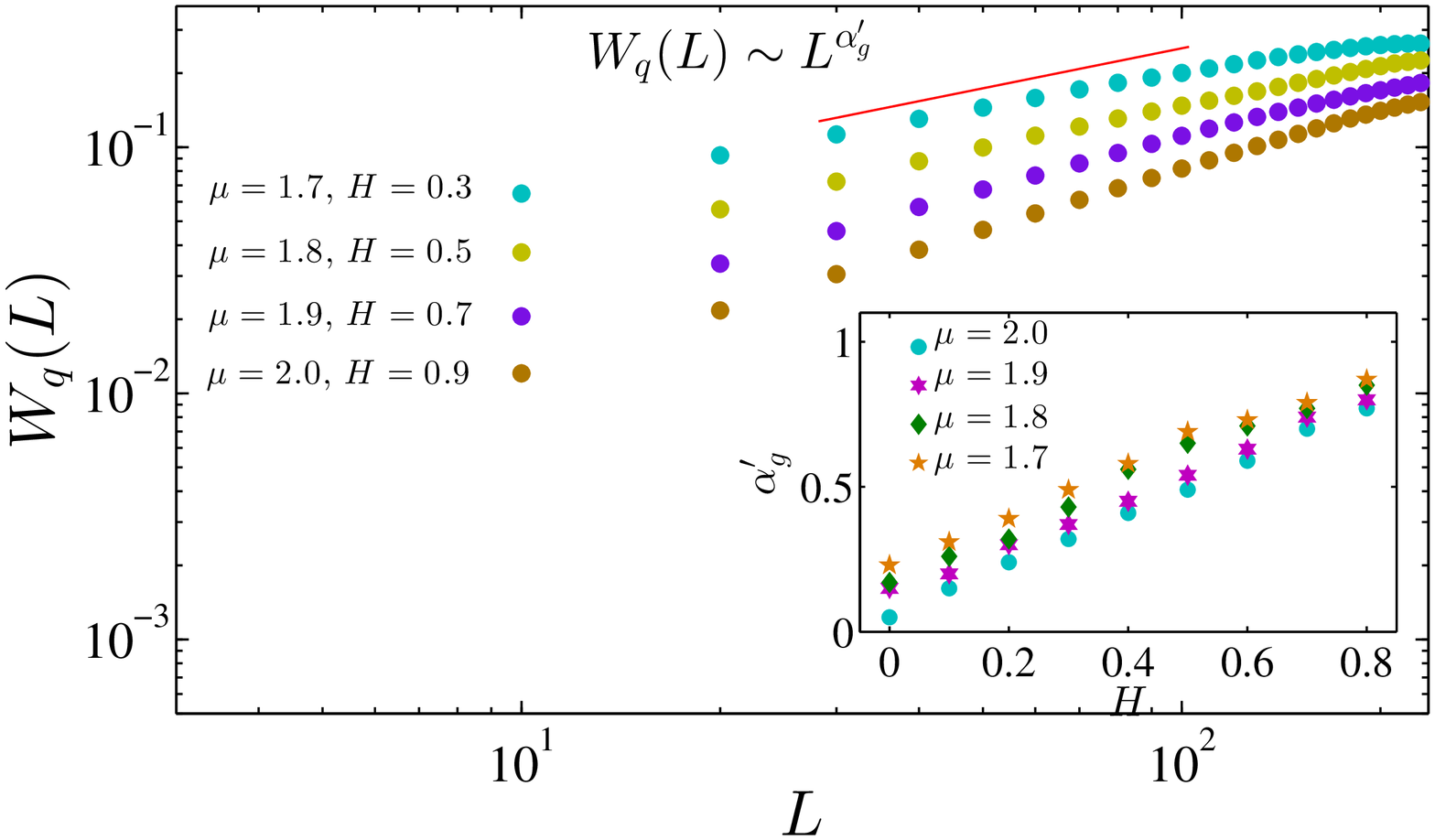}
\end{center}
\caption{(Color online) Top: Log-Log plot of $q$th order two-point correlation function $C_q(r)$ as a function of $r$ ($q=2/\mu$). The slope of this plot corresponds to local roughness exponent $\alpha^\prime_l$ (Inset). Bottom: Log-Log plot of the $q$th order profile width
$W_q(L)$ with respect to the window size $L$ ($q=2/\mu$). The slope of this plot corresponds to global roughness exponent $\alpha^\prime_g$ (Inset).   
}
\label{figur4}
\end{figure} 

By studying the scaling properties of the probability distribution function $\mathcal{P}\lbrace \mathcal{L} \rbrace$ of the $\mu$-stable fractional L\'evy motion, we obtain an interesting result on the existence of the scaling properties:
\begin{eqnarray}
p(x) \sim 1/x^{\mu^\prime},
\end{eqnarray}
with the tail scaling exponent $\mu^\prime$. Results from a simulation of $\mu$-stable fractional L\'evy motion are shown in Fig. \ref{figur3}. It is noteworthy that there is a linear relation between $\mu^\prime$ and the control parameters of the $\mu$-stable fLm model, $\mu$ and the \textit{Hurst} index $H$,. The relation is
\begin{eqnarray}
\mu^\prime \approx 1+\mu +H.
\end{eqnarray} 
To present additional evidences of the existence of scale invariant features of our proposed non-Gaussian random field ($\mu$-stable fLm), in the following, we will examine other properties of the model \textit{i.e.} $q$th order correlation function and the $q$th order global width.

\subsubsection{Local and global roughness exponents}

Two-point correlation function of the Gaussian random field configuration $h$, for the most statistical models, behaves like: 
$C(r) \equiv \langle \left[ h(\vec{x})-h(\vec{x}^\prime) \right]^2 \rangle^{1/2} \sim r ^{\alpha_l}$, 
where $r = |\vec{x}-\vec{x}^\prime|$, and the parameter $\alpha_l$ is called the local roughness exponent \cite{surface2}. The value of the scaling exponent $\alpha_l$ can be measured by a linear
fit $C(r)$, with $r$ in $\log-\log$ scale. Like the two-point function $C(r)$, the $q$th order correlation function $C_q(r)\equiv \langle \left[ h(\vec{x})-h(\vec{x}^\prime) \right]^q \rangle^{1/q} \sim r ^{\alpha_l}$ can also be used to analyze the self-affine properties of the random fluctuations. More specifically, we need to analyze the scaling properties of $C_q(r)$ for the random fluctuations with heavy-tailed distributions. In Fig. (\ref{figur4}), we have plotted the scaling relation between $q$th order two-point correlation function $C_q(r)\sim r ^{\alpha^\prime_l}$ for the $\mu$-stable fLm model. The insets of Fig. (\ref{figur4}) show $\alpha^\prime_l$ as function of $H$. 
We note that the local roughness exponent $\alpha^\prime_l$, as shown in the top panel of Fig. (\ref{figur4}), is a linear function of the \textit{Hurst} index $H$.

\begin{figure}[t]
\begin{center}
\includegraphics[width=8cm,clip]{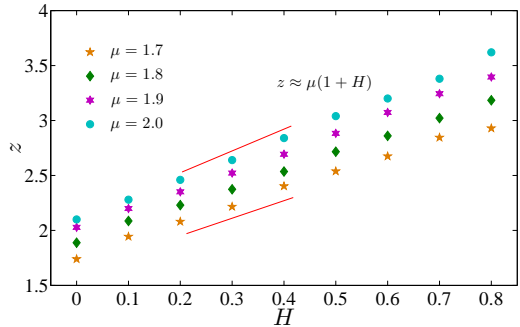}
\end{center}
\caption{(Color online) The relation between the fractional order $z$ and the \textit{Hurst} index $H$. Based on the numerical simulations, the $\mu$-stable fractional Le\'vy motion (with the stable parameter $\mu$ and the \textit{Hurst} index $H$) is a perfect candidate for the solution to the Eq. (\ref{ew mh}).
}
\label{figur5}
\end{figure} 

We also calculated the $q$th order global width of the fluctuations $W_q(l)\equiv \left\langle{\left[h(\vec{x}) - \bar{h} \right]^q}  \right\rangle^{1/q}$, from which the global roughness exponent $\alpha^\prime_g$ is extracted using the scaling relation $W_q(l)\sim l^{\alpha^\prime_g}$. In the bottom panel of Fig. (\ref{figur4}), we have presented the $q$th order global width for different values of the \textit{Hurst} index $H$ and stable parameter $\mu$. From these results, we extract the scaling exponent $\alpha^\prime_g$. In the inset of the same figure, we present $\alpha^\prime_g$ as a function of $H$. Conspicuously, the global roughness exponent $\alpha^\prime_g$ linearly depends on the parameter $H$. Our results for the exponents $\alpha^\prime_l$ and $\alpha^\prime_g$ (see Fig. (\ref{figur4})) are the same within statistical errors. This indicates that the $\mu$-stable fractional L\'evy motion obeys the mono-fractal random fields laws. It is worth mentioning that, for a given mono-fractal random field $\alpha_l=\alpha_g$ \cite{surface2}.

Here we have supposed that the solution to Eq. (\ref{fractional edwards wilkinson}) (in the limit $t \to \infty$) and the $\mu$-stable fractional L\'evy motions, are statistically equivalent. If we assume that the roughness exponent $\alpha = \frac{z+(1-\mu)d}{\mu}$ for the fractional Langevin equation (\ref{fractional edwards wilkinson}) is similar to the one which has been measured for the $\mu$-stable fractional L\'evy motion, we are able to find a direct relation between the fractional order $z$ and the \textit{Husrt} index $H$. For two-dimensional Gaussian process, this relation is given by $z = 2H+2$.
As shown in Fig. (\ref{figur5}), we present our numerical estimates for the fractional order $z$ for the corresponding value of the parameter $H$. Our numerical test shows that one can approximately find a linear relation $z\approx \mu(1+H)$ for all the values of the stable parameter $\mu$.

In the next section, we will numerically calculate the scaling exponents related to the contour lines of the $\mu$-stable fractional L\'evy motion. 

\begin{figure}[t]
\begin{center}
\includegraphics[width=40mm]{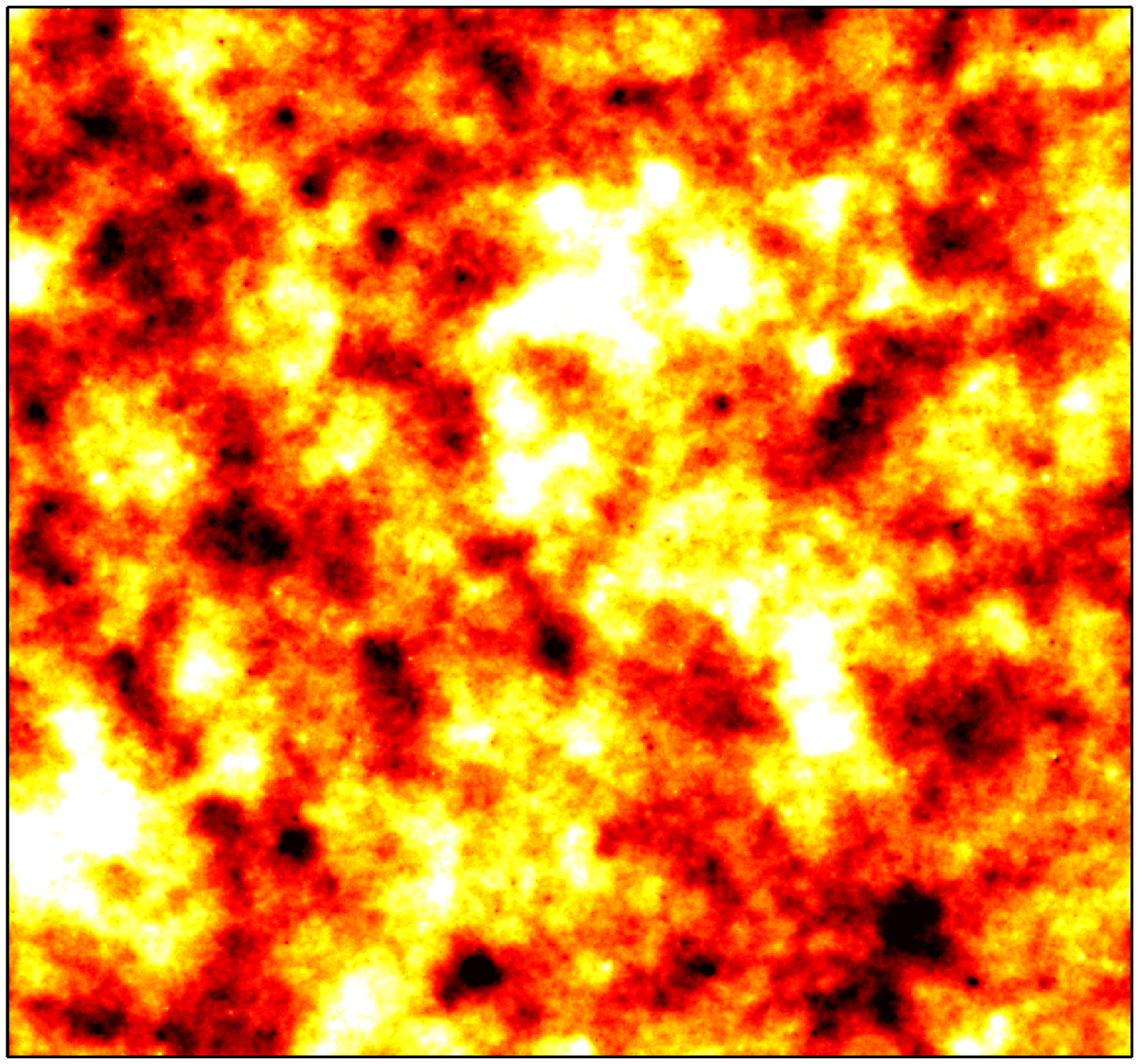}
\includegraphics[width=40mm]{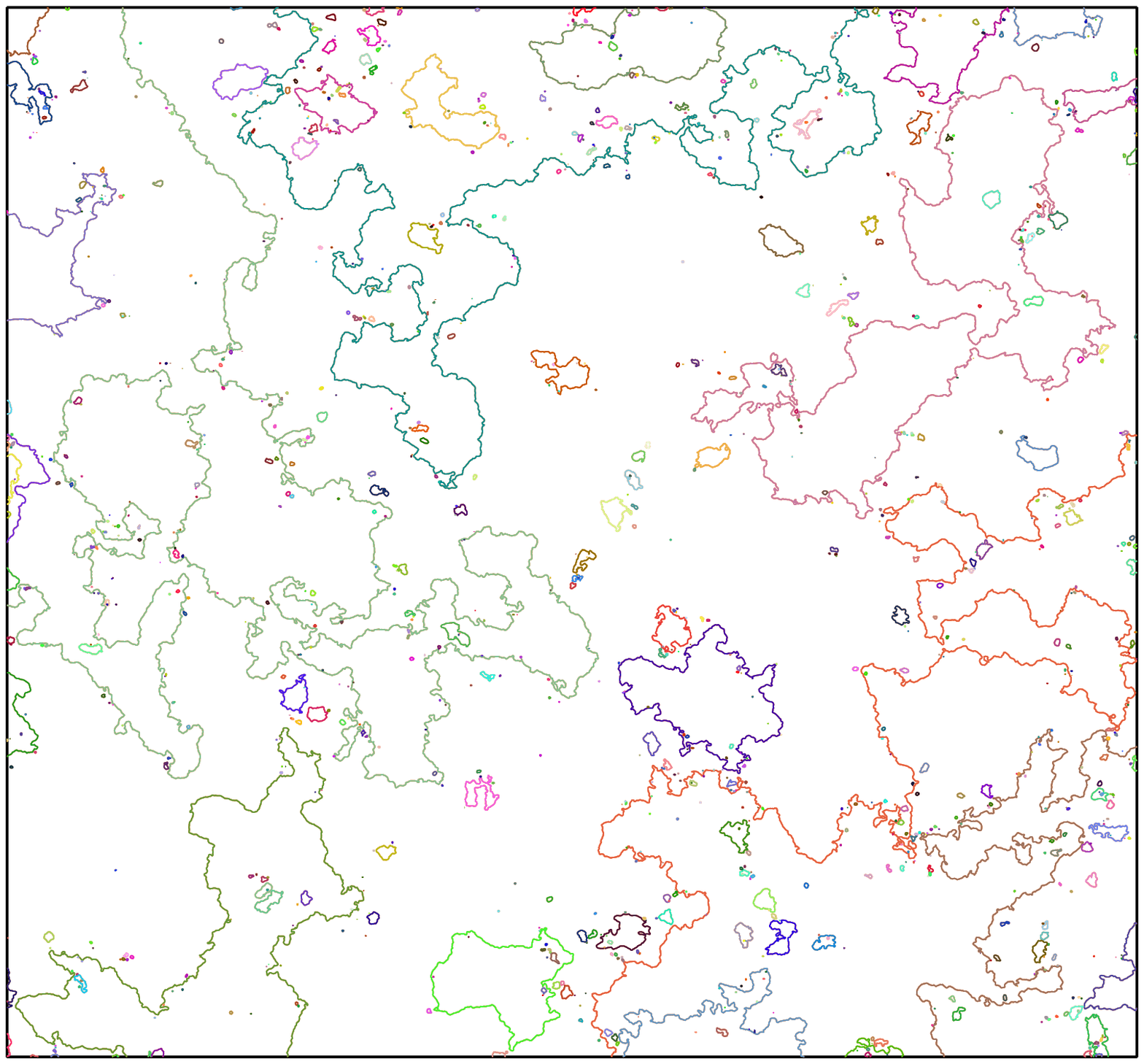}
\end{center}
\caption{(Color online) Color plot and contour lines of $\mu$-stable fractional L\'evy motion with different control parameters $\mu$ and $H$. The contour loop ensemble consists of closed non-intersecting loops that
connects points of $\bar{h} =: 0$: (top) $\mu = 1.4$ and $H = 0.0$, (middle) $\mu = 1.8$ and $H = 0.3$, (bottom) $\mu = 1.93$ and $H = 0.8$.}
\label{figur6}
\end{figure}

\begin{figure}[t]
\begin{center}
\includegraphics[width=8cm,clip]{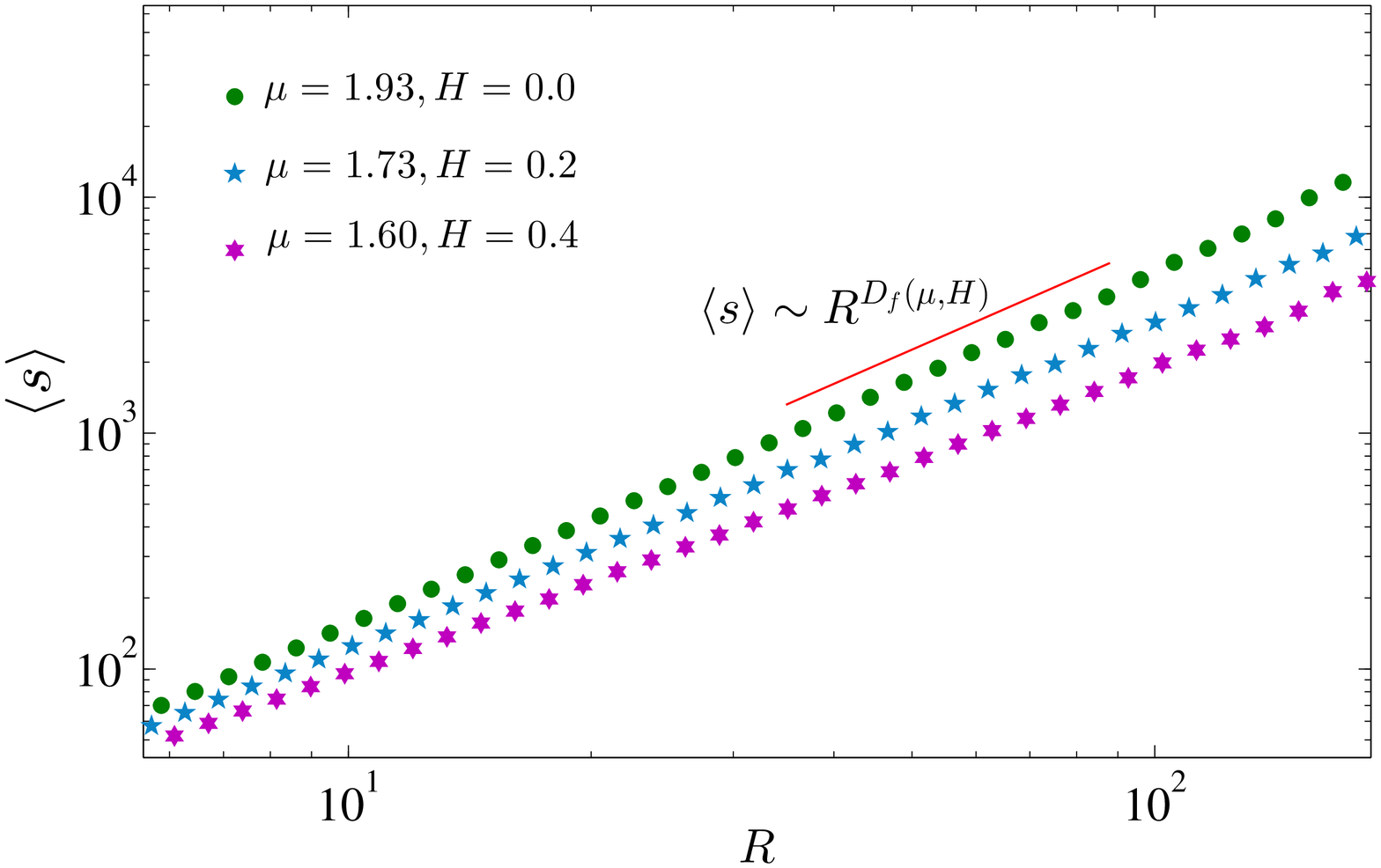}
\includegraphics[width=8cm,clip]{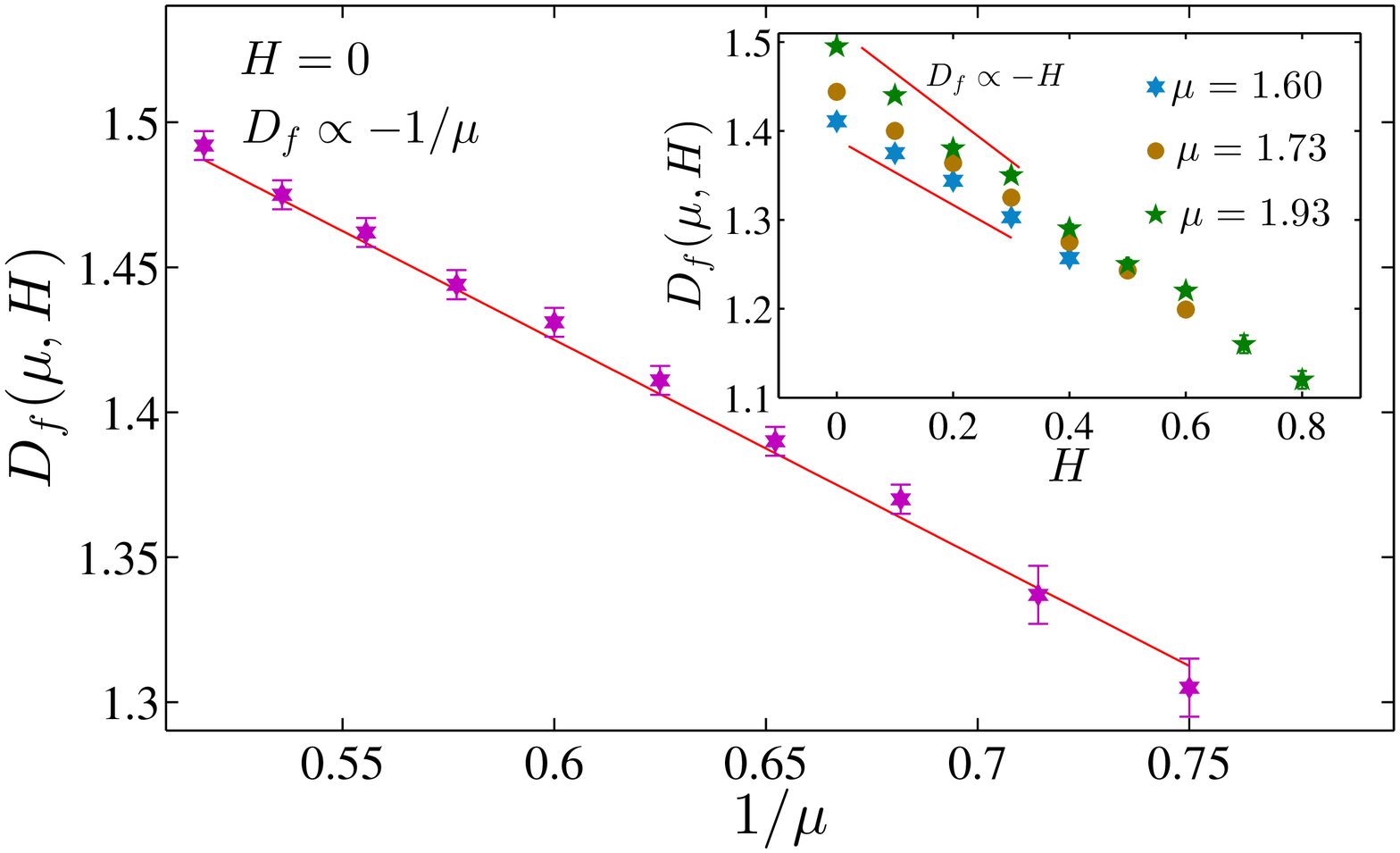}
\end{center}
\caption{(Color online) (Top) The power-law scaling relation between the mean value of the loop length $\langle s \rangle$ and the gyration radius $R$ for
contour lines of $\mu$-stable fractional L\'evy motion. (Bottom) The relation between fractal dimension $D_f(\mu,H)$ and the parameters $\mu$ and $H$.   
}
\label{figur7}
\end{figure}

\subsection{Scaling properties of contour loop ensembles}

Consider the random profile $h(\vec{x})\equiv h(x,y)$ of a scale invariant two-dimensional self-affine stochastic field that obeys the scaling law
 $h(\lambda \vec{x}) \simeq \lambda ^{\alpha} h(\vec{x})$. The iso-height lines of $h(x,y)$ at the level set $h(\vec{x}) = h_0$ contains many non-intersecting closed loops; in other words, a set of ``contour ensembles''. In Fig. (\ref{figur6}), we have plotted the contour loop ensembles for $\mu$-stable fLm with different values of the control parameters $\mu$ and $H$.
 
According to the scaling theory of random fields, a contour loop ensemble of a mono-fractal random Gaussian surface (governed by a Gaussian distribution Eq. (\ref{Gaussian fields PDF})) is a scale-invariant geometrical object \cite{kondevprl,kondevpre}. A key question about non-Gaussian random fields is whether there is any relation between the contour loop exponents and the roughness exponent $\alpha$ of the process or the control parameters $\mu$ and $H$.

One of the most interesting aspects of the scale invariant contour loop ensembles is a scaling relation between the mean loop length $\langle s \rangle$ and loop radius $R$, such that
\begin{eqnarray}\label{loop fractal dimesion}
\langle s \rangle \sim R^{D_f},
\end{eqnarray}
where the exponent $D_f$ is called the fractal dimension of contour loops. The radius $R$ for a given closed loop is defined by $R^2 = \frac{1}{N}\sum_{i=1}^N |\mathbf{r}_i - \mathbf{r}_c|^2$, where $\mathbf{r}_c=\frac{1}{N}\sum_{i=1}^N \mathbf{r}_i$ is the center of mass. It is worth mentioning that for a Gaussian random field, with roughness exponent $\alpha$,  
the fractal dimension $D_f$ follows from $D_f = \frac{3-\alpha}{2}$ \cite{kondevprl,kondevpre}.

In order to evaluate the fractal dimension of the contour loops $D_f$, we examined the scaling relation between the mean value of the loop perimeter $\langle s\rangle$ and $R$, according to Eq. (\ref{loop fractal dimesion}).
With regards to the contour loop ensemble of the non-Gaussian $\mu$-stable fractional L\'evy field, there is a scaling relation between $\langle s\rangle$ and $R$ with the fractal dimension exponent $D_f(\mu,H)$, as shown in Fig. (\ref{figur7}). The fractal dimension of the closed contours is measured using the linear fit in the scaling regime. The graph of the fractal dimension $D_f(\mu,H)$, as a function of $1/\mu$, (Fig (\ref{figur7})), shows a linear dependence $D_f\propto 1/\mu$. In addition, in the inset we demonstrate the fractal dimension $D_f(\mu,H)$ for a fixed value of the parameter $\mu$. The results of
simulations behave like $D_f \propto H$.  
Our studies are currently in progress to determine the exact
dependence of $D_f(\mu,H)$ on $\mu$ and $H$. It is important to note that, the relation $D_f =\frac{3-\alpha}{2} $ is not true for the $\mu$-stable fractional L\'evy field (except when $\mu = 2$).

\begin{figure}[t]
\begin{center}
\includegraphics[width=8cm,clip]{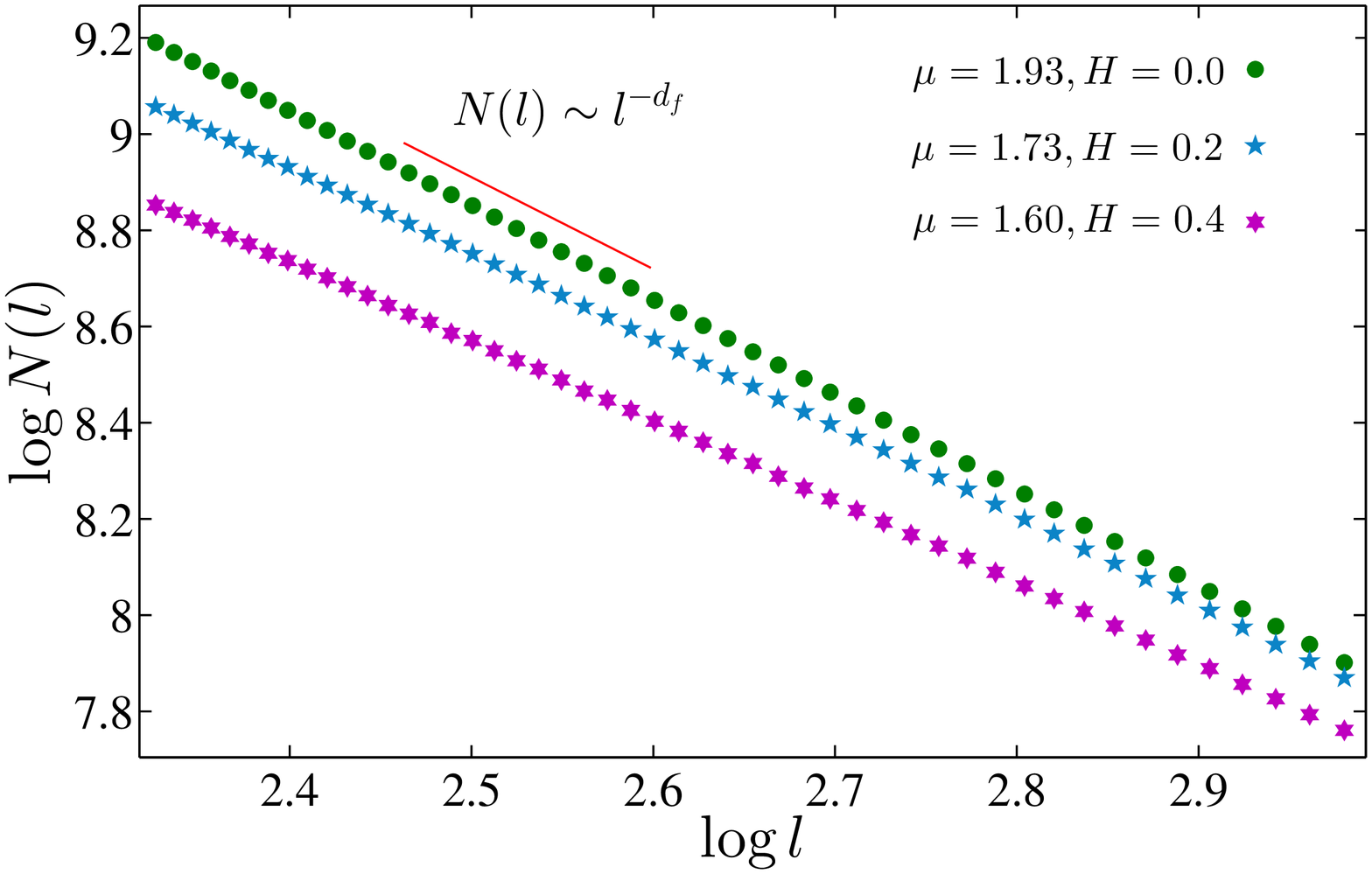}
\includegraphics[width=8cm,clip]{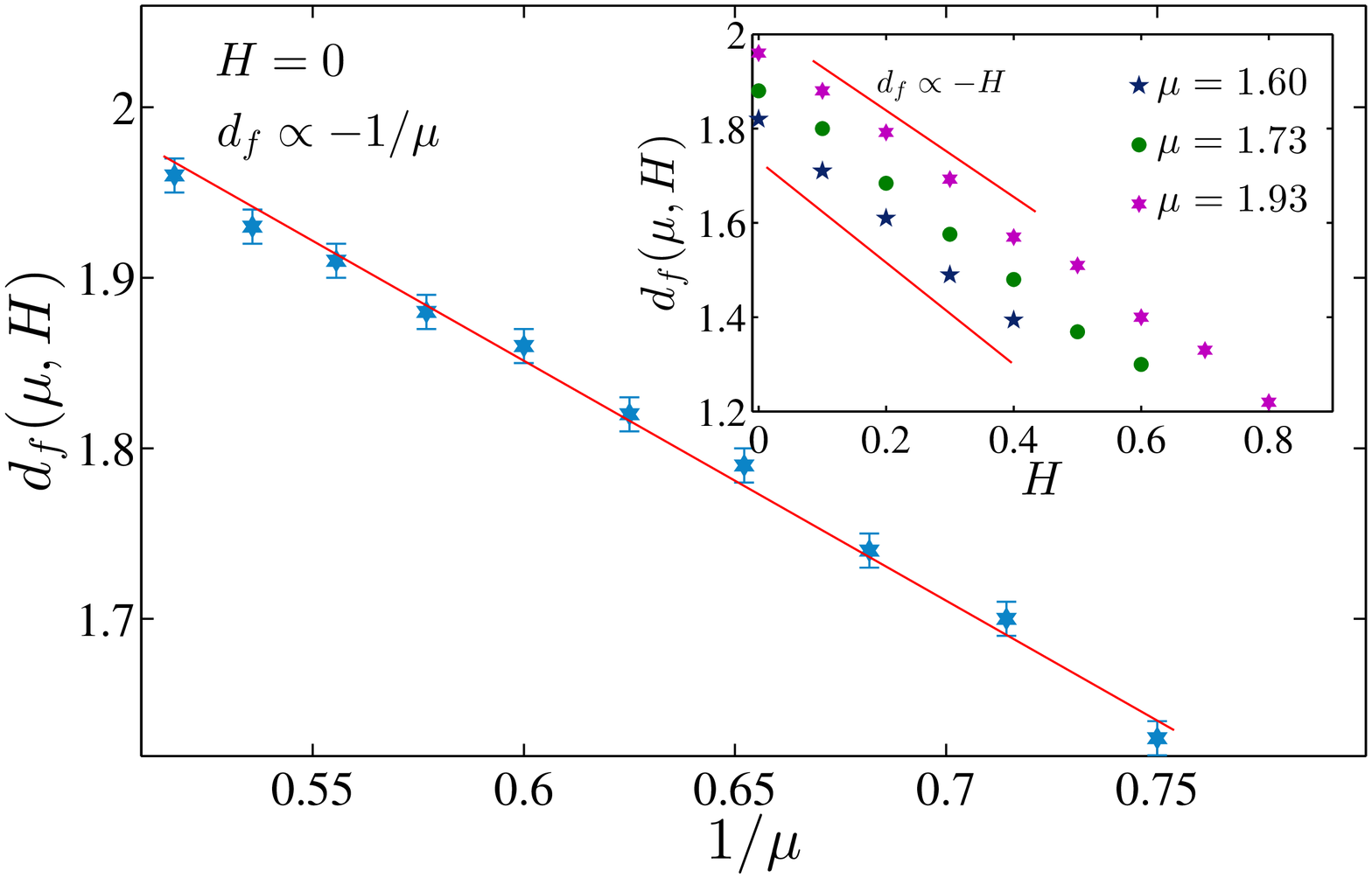}
\end{center}
\caption{(Color online) (Top) The power-law scaling relation between the number of boxes $N(l)$ with size $l$ needed to cover a given
contour set ensemble of $\mu$-stable fractional L\'evy motion. (Bottom) The relation between fractal dimension $d_f(\mu,H)$ and the parameters $\mu$ and $H$. 
}
\label{figur8}
\end{figure}

Another possible way for examining the self-similarity of the contour lines is box-counting approach \cite{mandelbrot}. For an
object with deterministic self-similarity, one can determine the fractal dimension $d_f$ based on the scaling relation $N(l)\sim l^{-d_f}$, where
$N(l)$ is the number of boxes with size $l$ needed for covering the fractal object. 
It is believed \cite{mandelbrot,kondevpre}, for a Gaussian random field with roughness exponent $\alpha$, that the fractal dimension of all contours is defined by $d_f = 2-\alpha$.
The scaling relation between $\log N(l)$ and $\log l$, for $\mu$-stable fractional L\'evy motion, is presented in Fig. (\ref{figur8}). The slope of this plot determines the fractal dimension $d_f(\mu,H)$. At the bottom of Fig. (\ref{figur8}) we have shown the measured value of $d_f(\mu,H)$. We emphasize that, interestingly, the relation $d_f = 2-\alpha$ is not valid for non-Gaussian power-law random fields.

\section{Conclusion}\label{section 5}

In this paper, we have presented a generalization of Edwards-Wilkinson equation (see Eq. (\ref{fractional edwards wilkinson})) to investigate the dynamics and statistical properties of correlated random fluctuations with heavy-tailed probability distribution function. Following the scaling analysis presented here, we calculated the roughness exponent $\alpha$ associated with the non-Gaussian fluctuations with a heavy-tailed PDF. The Eq. (\ref{fractional edwards wilkinson}) in the limit $t\to \infty$ has a solution for an arbitrary $z$ and $\mu$. For  example, the Gaussian free field corresponds to $z=2$ and $\mu=2$. Also the fractional Brownian motion, a well-known example of correlated Gaussian fluctuations, corresponds to $z\neq 2$ and $\mu=2$. For arbitrary values of $z$ and $0<\mu <2$ the solution to Eq. (\ref{fractional edwards wilkinson}) is expected to be a non-Gaussian correlated random field. 
Using the
$\mu$-stable fractional L\'evy motion, we proposed a way to analyze the statistical properties of the random correlated fluctuations with heavy-tailed PDF. One of the special features is
that the contour lines of the
introduced fluctuations in two spatial dimensions are self-similar. The fractal dimensions of $\mu$-stable fLm contours do not follow the famous behavior of the Gaussian random fluctuations, $D_f=\frac{3-\alpha}{2}$ and $d_f = 2-\alpha$\cite{kondevpre,kondevprl}, although some numerical and experimental observations have shown that the same relations work for non-Gaussian fluctuations \cite{hosseinabadi2012,hosseinabadi2014,najafi2016,cmbcontours}. The
key point to be gleaned from our analysis is that the scaling theory reported for Gaussian random fluctuations \cite{kondevprl,kondevpre} holds only for fast decaying distributions. Since most of the well-known statistical fluctuations exhibiting a heavy-tailed distribution, it seems that we need also the scaling theory for those random fluctuations with heavy tailed PDFs. Specifically, we need a better understanding of the scaling properties of the contour loop ensembles in two-dimensional non-Gaussian random fields with heavy-tailed PDFs. It is quite natural to numerically find  some other scaling exponents associated with contour lines and find other relevant aspects of random fluctuations with heavy-tailed PDF. 

\textbf{Acknowledgments:}  
The authors would like to thank the anonymous referees for their constructive comments that greatly contributed to improving this paper. MGN thanks M. R. Nematollahi for reading the manuscript and M. Nattagh-Najafi for helpful discussions.


\begin{thebibliography}{99}
\bibitem{ps1} For $1 < \mu < 2$ the second moment diverges but $q$th moment $\langle |\eta_\mu|^q \rangle$ of the random variable $\eta_\mu$
exists and is finite only if $0 < q < \mu$. In the interval $0 < \mu < 1$ the first moment does not exist \cite{stochastic book,gardiner}.
\end{thebibliography}

\end{document}